\documentclass[pra, twocolumn, floatfix,superscriptaddress]{revtex4}
\usepackage{graphicx}
\usepackage{epstopdf}
\usepackage{amsmath, amsfonts, amssymb, bm,color}
\def\mpik{Max-Planck-Institut f\"ur Kernphysik, Saupfercheckweg 1, D-69117
Heidelberg, Germany}
\def\ifa{Institute of Applied Physics, Academiei str. 5, MD-2028 Chi\c{s}in\u{a}u, Moldova}
\begin{document}
%%%%%%%%%%%%%%%%%%%%%%%%%%%%%%%%%%%%%%%%%%%%%%%%%%%%%%%%%%%%%%%%%
\title{Spontaneous decay processes in a classical strong low-frequency laser field}
%%%%%%%%%%%%%%%%%%%%%%%%%%%%%%%%%%%%%%%%%%%%%%%%%%%%%%%%%%%%%%%%%
\author{Mihai  \surname{Macovei}}
\email{macovei@phys.asm.md}
\affiliation{\mpik}
\affiliation{\ifa}

\author{J\"{o}rg \surname{Evers}}
\email{joerg.evers@mpi-hd.mpg.de}
\affiliation{\mpik}

\author{Christoph H. \surname{Keitel}}
\email{keitel@mpi-hd.mpg.de}
\affiliation{\mpik}

\date{\today}
%%%%%%%%%%%%%%%%%%%%%%%%%%%%%%%%%%%%%%%%%%%%%%%%%%%%%%%%%%%%%%%%%
\begin{abstract}
The spontaneous emission of an excited two-level emitter driven by a strong classical coherent low-frequency  electromagnetic field is investigated. 
We find that for relatively strong laser driving, multi-photon processes are induced, thereby opening additional decay channels for the atom. We 
analyze the interplay between the strong low-frequency driving and the interfering multiphoton decay channels, and discuss its implications for the 
spontaneous emission dynamics. 
\end{abstract}
%%%%%%%%%%%%%%%%%%%%%%%%%%%%%%%%%%%%%%%%%%%%%%%%%%%%%%%%%%%%%%%%%%%
%\pacs{42.50.Hz, 42.50.Ct, 42.25.Hz}
\maketitle
%%%%%%%%%%%%%%%%%%%%%%%%%%%%%%%%%%%%%%%%%%%%%%%%%%%%%%%%%%%%%%%%%%%
%%%%%%%%%%%%%%%%%%%%%%%%%%%%%%%%%%%%%%%%%%%%%%%%%%%%%%%%%%%%%%%%%%%
\section{Introduction}
%%%%%%%%%%%%%%%%%%%%%%%%%%%%%%%%%%%%%%%%%%%%%%%%%%%%%%%%%%%%%%%%%%%
Spontaneous emission (SE) is a basic process occurring in excited quantum systems coupled to environments~\cite{se1,se2,se3,ag,al}. Since it typically 
competes with coherent processes induced, e.g., by laser fields, its manipulation or even control is of vital importance for many applications. The SE 
rate of an atom depends on the transition dipole moment and the density of states of the environment~\cite{se3,ag,al}. Therefore, 
a first control approach is to suitably modify the environment's density of states, e.g., using cavities~\cite{kleppner,purcel,exp-cav1,exp-cav2,exp-cav3} 
or photonic crystals~\cite{yabl,john,pcexp,review-pc2,review-pc}. 
An alternative approach is to control the coupling between atom and environment, which typically involves atomic coherence and quantum interference 
effects~\cite{fc,rew,sczb,gxl}. 
% modulation
For example,  slow or fast transition-frequency modulations~\cite{parr,pasp,wys} were shown to allow for substantial suppression of the SE of an excited 
two-level emitter inside a leaking cavity~\cite{mod1,mod2,mod3}, and such control schemes can be extended to dc fields~\cite{agw_p}. The possible 
effects of external modulations or perturbations on the SE into potentially structured environments can also be classified  on a more general level~\cite{shap,kur}. 
% SGC
Another ansatz to control SE facilitates spontaneously generated coherences (SGC)~\cite{ag,fc,rew}, which may suppress the SE of particular excited 
states via destructive interference of different decay pathways. Based on this, a broad variety of applications has been proposed, including lasing 
without inversion~\cite{lwi1,lwi2,lwi3} and the stabilization of coherences in quantum computing~\cite{stab1,stab2}, and SGC have also been observed 
experimentally~\cite{sgc-1, sgc-2}.
% Review
Related approaches to control SE are reviewed in~\cite{review-pc2,review-pc,fc,rew}.

A further ansatz to modify and substantially slow-down the usual spontaneous decay of excited atoms was proposed in~\cite{ek1,ek2}, based on the 
application of a strong  low-frequency electromagnetic field (LFF) to the excited emitter. A perturbative analysis in the LFF-atom coupling showed that 
the LFF induces additional multiphoton decay pathways, in which the atom exchanges photons with the field during the spontaneous decay. These arise 
since the model includes off-resonant excited auxiliary energy levels as possible intermediate states in the multiphoton processes,  in addition to the two 
energy levels involved in the natural spontaneous decay. Importantly, ``low frequency'' here refers to driving fields with frequency lower than the 
spontaneous emission line width, such that the multiphoton pathways are indistinguishable and may interfere, thereby affecting the usual spontaneous 
decay. The LFF-induced multiphoton pathways  were interpreted in~\cite{ek1,ek2} in terms of an effective upper-state multiplet of energy levels, with 
the spontaneous-emission modification arising from the interference of the decay amplitudes out of the different multiplet states. However, the initial 
work triggered further discussions~\cite{com1,com2,com3}, in which in particular the role of the multiphoton pathways and the interpretation in terms 
of an excited-state multiplet was questioned~\cite{com2}. This invites further investigations on the effect of field-induced multiphoton processes on 
spontaneous emission and their interpretation.

Motivated by this, here, we investigate the spontaneous emission of an excited two-level quantum emitter interacting with an intense classical LFF. 
Unlike in the previous work~\cite{ek1,ek2}, we restrict the analysis to a two-level system, and thereby explicitly exclude the possibility to induce 
interfering multi-photon pathways involving  off-resonant auxiliary states. This choice  allows us to explore the significance of these processes, but 
also enables the calculation of higher-order effects in the LFF-atom interaction.
We show that despite the absence of intermediate states, the strong LFF still may induce interfering multi-photon evolution. However, the nature of 
these pathways is very different. They proceed directly from the excited to the ground state, but involve the interaction of the atom  with different 
harmonics of the LFF, which again can be interpreted as the exchange of different numbers of photons of the LFF throughout the atomic transition. 
As a result of these different pathways, we again find that strong LFF driving may modify the standard exponential spontaneous decay law, 
either slowing down or accelerating the decay. However, the effect is not as pronounced as that predicted in \cite{ek1,ek2}. This suggests 
that the additional decay pathways via off-resonant auxiliary states are crucial.

We note that it is well-known  in general that intense  electromagnetic fields may substantially modify the atomic dynamics
~\cite{jetp,reiss,reiss1,milonni,chk,rev-intenseII,rev-intense}. 
Relevant to the SE control via multiphoton pathway interference  discussed here, the probabilities for relevant multi-photon transitions in two-level systems 
interacting with a strong and coherent classical electromagnetic field of frequency much lower than the involved transition frequencies were 
calculated~\cite{jetp1}, as well as related light emission and absorption processes~\cite{jetp2}, and the multi-photon resonance-induced fluorescence 
of strongly driven two-level systems under frequency modulation~\cite{art5,fmod,art6}. It was also shown that various superposition states may occur 
via multi-photon resonant excitations in hydrogen-like atoms~\cite{hamlet}, and  methods were developed to deal with the laser-dressing of the atoms~\cite{dress}, 
or to calculate the relevant transition elements~\cite{sal}. However, in the above-mentioned works on the quantum dynamics of isolated two-level systems 
interacting with a low-frequency and strong classical electromagnetic field, ``low-frequency'' typically refers to field frequencies much lower than the 
involved transition frequencies, but not than the spontaneous emission linewidths. Also, these works do not investigate explicitly the spontaneous decay.

%%%%%%%%%%%%%%%%%%%%%%%%%%%%%%%%%%%%%%%%%%%%%%%%%%%%%%%%%%%%%%%%%%%
\section{Analytical framework}
%%%%%%%%%%%%%%%%%%%%%%%%%%%%%%%%%%%%%%%%%%%%%%%%%%%%%%%%%%%%%%%%%%%
The Hamiltonian of a two-level emitter interacting with a strong low-frequency field of frequency $\omega$ as well as with the environmental vacuum 
modes of the electromagnetic field reservoir is:
%%%%%%%%%%%%%%%%%%%%%%%%%%%%%%%%%%%%%%%%%%%%%%%%%%%%%%%%%%%%%%%%%%
\begin{eqnarray}
H&=&\sum_{k}\hbar\omega_{k}a^{\dagger}_{k}a_{k} + \hbar \omega_{0}S_{z} - \hbar \Omega \cos(\omega t + \phi)(S^{+}+S^{-}) \nonumber \\
&+& i \sum_{k}(\vec g_{k}\cdot \vec d)(a^{\dagger}_{k}-a_{k})(S^{+}+S^{-}). 
\label{Ham}
\end{eqnarray}
%%%%%%%%%%%%%%%%%%%%%%%%%%%%%%%%%%%%%%%%%%%%%%%%%%%%%%%%%%%%%%%%%%
Here, $\omega_{0}$ is the transition frequency among the involved states $|2\rangle \leftrightarrow |1\rangle$ with the transition dipole $d$, whereas 
$\Omega$ is the corresponding Rabi frequency and $\phi$ is the laser absolute phase. The atom-vacuum coupling strength is 
$\vec g_{k}=\sqrt{2\pi\hbar\omega_{k}/V}\vec e_{\lambda}$ where $V$ is quantization volume while $\vec e_{\lambda}$ is the photon polarization vector 
with ${\lambda = 1,2}$. $a^{\dagger}_{k}$ and $a_{k}$ are the creation and annihilation operators for the photons with the momentum $\hbar k$, energy 
$\hbar \omega_{k}$ and polarization $\lambda$ satisfying standard commutation relations for bosons. Further, $S^{+}=|2\rangle \langle 1|$, 
$S^{-}=[S^{+}]^{\dagger}$ and $S_{z}= (|2\rangle\langle 2| - |1\rangle\langle 1|)/2$ are the well-known quasi-spin operators obeying the commutation 
relations for SU(2) algebra. In the Hamiltonian (\ref{Ham}) the first three components are, respectively, the free energies of the environmental electromagnetic 
vacuum modes and atomic subsystems together with the laser-atom interaction Hamiltonian. The last term accounts for the interaction of a two-level emitter 
with the surrounding electromagnetic field vacuum modes.

The quantum dynamics of any atomic operator $Q$ is determined by the Heisenberg equation 
%%%%%%%%%%%%%%%%%%%%%%%%%%%%%%%%%%%%%%%%%%%%%%%%%%%%%%%%%%%%%%%%%%
\begin{eqnarray}
\frac{d}{dt}Q(t) = \frac{i}{\hbar}\bigl[H,Q\bigr]. \label{qq}
\end{eqnarray} 
%%%%%%%%%%%%%%%%%%%%%%%%%%%%%%%%%%%%%%%%%%%%%%%%%%%%%%%%%%%%%%%%%%
In the following, we perform a spin rotation \cite{pasp,hamlet}, $U(t)=\exp\bigl[2i\theta(t)S_{y}\bigr]$, to the entire Hamiltonian which transforms it as 
follows 
%%%%%%%%%%%%%%%%%%%%%%%%%%%%%%%%%%%%%%%%%%%%%%%%%%%%%%%%%%%%%%%%%%
\begin{eqnarray}
\bar H = UHU^{-1}-2\bigl(d\theta(t)/dt\bigr)US_{y}U^{-1}. \label{hb}
\end{eqnarray} 
%%%%%%%%%%%%%%%%%%%%%%%%%%%%%%%%%%%%%%%%%%%%%%%%%%%%%%%%%%%%%%%%%%
Here, $\theta(t) \equiv \theta = \arctan\bigl[\bigl(2\Omega/\omega_{0}\bigr)\cos(\omega t + \phi)\bigr]/2$,
while $S_{y}=(S^{+} - S^{-})/(2i)$. Then, the total Hamiltonian reads as follows:
%%%%%%%%%%%%%%%%%%%%%%%%%%%%%%%%%%%%%%%%%%%%%%%%%%%%%%%%%%%%%%%%%%
\begin{eqnarray}
\bar H &=& \sum_{k}\hbar\omega_{k}a^{\dagger}_{k}a_{k} + 2\hbar \bar \Omega(t)R_{z} + i\hbar \alpha(t)(R^{-}-R^{+}) \nonumber \\ 
&+& i\sum_{k}(\vec g_{k}\cdot \vec d)(a^{\dagger}_{k} - a_{k})\bigl(\cos{2\theta}(R^{+} + R^{-}) \nonumber \\
&-& 2\sin{2\theta}R_{z}\bigr), \label{barH}
\end{eqnarray}
%%%%%%%%%%%%%%%%%%%%%%%%%%%%%%%%%%%%%%%%%%%%%%%%%%%%%%%%%%%%%%%%%%
where 
%%%%%%%%%%%%%%%%%%%%%%%%%%%%%%%%%%%%%%%%%%%%%%%%%%%%%%%%%%%%%%%%%%
\begin{eqnarray}
\bar \Omega(t)=\sqrt{\bigl(\omega_{0}/2\bigr)^{2}+\Omega^{2}\cos^{2}\bigl(\omega t +\phi\bigr)}, \label{omb}
\end{eqnarray}
%%%%%%%%%%%%%%%%%%%%%%%%%%%%%%%%%%%%%%%%%%%%%%%%%%%%%%%%%%%%%%%%%%
whereas $\alpha(t) = \bigl(\omega/2\bigr)\Omega\cos\bigl(2\theta\bigr)\sin\bigl(\omega t + \phi\bigr)/\bar \Omega(t)$ and $\omega/\omega_{0} \ll 1$. 

The new quasi-spin operators, i.e. $R_{z}$ and $R^{\pm}$, can be represented via the old ones in the following way
%%%%%%%%%%%%%%%%%%%%%%%%%%%%%%%%%%%%%%%%%%%%%%%%%%%%%%%%%%%%%%%%%%
\begin{eqnarray}
R_{z} &=&S_{z}\cos{2\theta} - (S^{+}+S^{-})\sin{2\theta}/2, \nonumber \\
R^{+} &=&S^{+}\cos^{2}{\theta} - S^{-}\sin^{2}{\theta} + S_{z}\sin{2\theta}, \nonumber \\
R^{-} &=& [R^{+}]^{\dagger}, \label{rs}
\end{eqnarray}
%%%%%%%%%%%%%%%%%%%%%%%%%%%%%%%%%%%%%%%%%%%%%%%%%%%%%%%%%%%%%%%%%%
and obey the commutation relations: $[R^{+},R^{-}]=2R_{z}$ and $[R_{z},R^{\pm}]=\pm R^{\pm}$, similarly to the old-basis ones. 
The Hamiltonian (\ref{barH}), based on the unitary transformation $U(t)$, will allow us to follow the quantum dynamics of the excited 
two-level emitter where absorption of the external low-frequency field photons is incorporated naturally. This is not evident if one starts 
directly with the Hamiltonian (\ref{Ham}).

In what follows, we are interesting in laser-atom interaction regimes such  that $2\Omega/\omega_{0} < 1$. On the other side, the Rabi 
frequency $\Omega$ can be smaller, of the same order, or larger than the laser frequency $\omega$, respectively. Consequently, we 
expand the generalized Rabi frequency $\bar \Omega(t)$, in Exp.~(\ref{omb}), up to second order in the small parameter $2\Omega/\omega_{0}$, 
namely, 
%%%%%%%%%%%%%%%%%%%%%%%%%%%%%%%%%%%%%%%%%%%%%%%%%%%%%%%%%%%%%%%%%%
\begin{eqnarray}
\bar \Omega(t) \approx \frac{\omega_{0}}{2}\biggl (1 + \Omega^{2}/\omega^{2}_{0} + \Omega^{2}\cos{[2(\omega t +\phi)}]/\omega^{2}_{0} \biggr ). 
\label{grabi}
\end{eqnarray}
%%%%%%%%%%%%%%%%%%%%%%%%%%%%%%%%%%%%%%%%%%%%%%%%%%%%%%%%%%%%%%%%%%
Next, in the Hamiltonian (\ref{barH}), we pass to the interaction picture using the operator 
%%%%%%%%%%%%%%%%%%%%%%%%%%%%%%%%%%%%%%%%%%%%%%%%%%%%%%%%%%%%%%%%%%
\begin{eqnarray*}
V(t) = \exp\biggl[2i\int^{t}_{0}dt^{'}\bar \Omega(t^{'})R_{z}\biggr],
\end{eqnarray*}
%%%%%%%%%%%%%%%%%%%%%%%%%%%%%%%%%%%%%%%%%%%%%%%%%%%%%%%%%%%%%%%%%%
with Exp.~(\ref{grabi}), and write down the formal solution of the Heisenberg equation for the field operator $a^{\dagger}_{k}(t)$, 
$a_{k}(t)=[a^{\dagger}_{k}(t)]^{\dagger}$, that is,
%%%%%%%%%%%%%%%%%%%%%%%%%%%%%%%%%%%%%%%%%%%%%%%%%%%%%%%%%%%%%%%%%%
\begin{widetext}
\begin{eqnarray}
a^{\dagger}_{k}(t) &=& a^{\dagger}_{k}(0)e^{i\omega_{k}t} + \frac{(\vec g_{k}\cdot \vec d)}{\hbar}\int^{t}_{0}dt^{'}e^{i\omega_{k}(t-t^{'})}
\biggl \{ \sum^{\infty}_{m=-\infty}J_{m}(\eta) \biggl (R^{+}(t^{'})e^{i(\bar \omega_{0}t^{'}-\eta\sin{2\phi})}e^{2im(\omega t^{'}+\phi)} + H.c.\biggr)
\cos{2\theta} \nonumber \\
&-&2\sin{2\theta}R_{z}(t^{'})\biggr \}, 
\label{sol_ak}
\end{eqnarray}
\end{widetext}
%%%%%%%%%%%%%%%%%%%%%%%%%%%%%%%%%%%%%%%%%%%%%%%%%%%%%%%%%%%%%%%%%%
where 
\begin{eqnarray*}
\cos{2\theta} &\approx& 1 - \bigl(2\Omega/\omega_{0}\bigr)^{2}\cos^{2}\bigl(\omega t^{'} + \phi\bigr)/2, \\
\sin{2\theta} &\approx& \bigl(2\Omega/\omega_{0}\bigr)\cos\bigl(\omega t^{'} + \phi\bigr), 
\end{eqnarray*}
%%%%%%%%%%%%%%%%%%%%%%%%%%%%%%%%%%%%%%%%%%%%%%%%%%%%%%%%%%%%%%%%%%
and 
%%%%%%%%%%%%%%%%%%%%%%%%%%%%%%%%%%%%%%%%%%%%%%%%%%%%%%%%%%%%%%%%%%
\begin{eqnarray}
\bar \omega_{0}=\omega_{0}\bigl(1+\Omega^{2}/\omega^{2}_{0}\bigr). \label{bom}
\end{eqnarray}
%%%%%%%%%%%%%%%%%%%%%%%%%%%%%%%%%%%%%%%%%%%%%%%%%%%%%%%%%%%%%%%%%%
Here, we used the expansion via the $m$th-order Bessel function of the first kind, i.e., 
\begin{eqnarray*}
e^{\pm i\eta \sin(2\omega t + 2\phi)}=\sum^{\infty}_{m=-\infty}J_{m}(\eta)e^{\pm 2i m(\omega t +\phi)},
\end{eqnarray*}
with $J_{m}(\eta)$ being the corresponding ordinary Bessel function, whereas 
%%%%%%%%%%%%%%%%%%%%%%%%%%%%%%%%%%%%%%%%%%%%%%%%%%%%%%%%%%%%%%%%%%
\begin{eqnarray}
\eta = \frac{\Omega^{2}}{2\omega\omega_{0}}, \label{eta}
\end{eqnarray}
%%%%%%%%%%%%%%%%%%%%%%%%%%%%%%%%%%%%%%%%%%%%%%%%%%%%%%%%%%%%%%%%%%
stands as a control parameter. In the Markov approximation, we identify the following emission processes based on Exp.~(\ref{sol_ak}):
%%%%%%%%%%%%%%%%%%%%%%%%%%%%%%%%%%%%%%%%%%%%%%%%%%%%%%%%%%%%%%%%%%
\begin{eqnarray}
\int^{\infty}_{0}d\tau e^{i(\omega_{k} \mp \bar \omega_{0} \mp 2m\omega)\tau} &=& \pi \delta(\omega_{k} 
\mp \bar \omega_{0} \mp 2m\omega) \nonumber \\
&+&iP_{c}\frac{1}{\omega_{k} \mp \bar \omega_{0} \mp 2m\omega}, \nonumber
\end{eqnarray}
\begin{eqnarray}
\int^{\infty}_{0}d\tau e^{i(\omega_{k} \pm \omega)\tau} &=& \pi\delta(\omega_{k} \pm \omega) + iP_{c}\frac{1}{\omega_{k} \pm \omega}, 
\nonumber \\ \label{dgr}
\end{eqnarray}
%%%%%%%%%%%%%%%%%%%%%%%%%%%%%%%%%%%%%%%%%%%%%%%%%%%%%%%%%%%%%%%%%%
where $P_{c}$ is the Cauchy principal part. One can observe here that the spontaneous emission processes involve an even laser photon number, i.e., 
the emission occurs at frequencies: $\omega_{k} = \bar \omega_{0} \pm 2m\omega$, or $\omega_{k} = 2m\omega - \bar \omega_{0} >0$. This also 
means that the pumping field opens additional spontaneous decay channels that may interfere. Actually, the latter emission process implies that the sum 
frequency of the multiple absorbed photons is larger than the transition frequency - a situation not considered here. Apart from these processes there 
are also spontaneous transitions around the laser frequency $\omega$,  i.e, an induced laser photon absorption is followed by a spontaneously 
re-scattered photon of the same frequency. Thus, the whole quantum dynamics is influenced by the above mentioned processes. Notice the 
modification of the transition frequency due to the external low-frequency strong coherent electromagnetic pumping field, see expression~(\ref{bom}). 
Also, the contribution of $P_{c}$ leading to a small frequency Lamb shift compared to the one due to direct photon absorption is ignored here.

The solution (\ref{sol_ak}) has to be introduced in the Heisenberg equation for the mean value of any atomic subsystem's operators $Q$, namely,
%%%%%%%%%%%%%%%%%%%%%%%%%%%%%%%%%%%%%%%%%%%%%%%%%%%%%%%%%%%%%%%%%%
\begin{widetext}
\begin{eqnarray}
\frac{d}{dt}\bigl\langle Q(t)\bigr\rangle &-& \frac{i}{\hbar}\bigl\langle \bigl[\bar H_{0}, Q(t)\bigr]\bigr\rangle 
= \sum_{k}\frac{(\vec g_{k}\cdot \vec d)}{\hbar}\bigl\langle a^{\dagger}_{k}\bigl[2\sin{2\theta}R_{z} - 
\cos{2\theta}\sum^{\infty}_{n=-\infty}J_{n}(\eta)\bigl(R^{+}e^{i(\bar \omega_{0}t-\eta\sin{2\phi})}e^{2in(\omega t + \phi)} \nonumber \\
&+& H.c. \bigr), Q(t)\bigr]\bigr\rangle + H.c., 
\label{MeqD}
\end{eqnarray}
\end{widetext}
%%%%%%%%%%%%%%%%%%%%%%%%%%%%%%%%%%%%%%%%%%%%%%%%%%%%%%%%%%%%%%%%%%
where, in general, for the non-Hermitian atomic operators $Q$, the $H.c.$ terms should be evaluated without conjugating $Q$, i.e., by replacing $Q^{+}$
with $Q$ in the Hermitian conjugate part. The notation $\langle \cdots \rangle$ indicates averaging over the initial state of both the atoms and the vacuum 
environmental system, respectively. In the master equation (\ref{MeqD}), the Hamiltonian describing the coherent evolution of the qubit during multiple 
photon absorption/emission processes is given by 
%%%%%%%%%%%%%%%%%%%%%%%%%%%%%%%%%%%%%%%%%%%%%%%%%%%%%%%%%%%%%%%%%%
\begin{eqnarray}
\bar H_{0} &=& i\hbar \alpha(t)\sum^{\infty}_{n=-\infty}J_{n}(\eta)R^{-}e^{-i(\bar \omega_{0}t-\eta\sin{2\phi})}
e^{-2in(\omega t + \phi)} \nonumber \\
&+& H.c., \label{H0} 
\end{eqnarray}
%%%%%%%%%%%%%%%%%%%%%%%%%%%%%%%%%%%%%%%%%%%%%%%%%%%%%%%%%%%%%%%%%%
with $\alpha(t) \approx (\omega\Omega/\omega_{0})\sin(\omega t + \phi)$. Contrary to spontaneous emission processes, the coherent evolution 
involves an odd laser photon number, i.e., resonances occur when $\bar \omega_{0} + (2n \pm 1)\omega =0$, see also \cite{pop}. The final expression 
for the master equation in the Born-Markov approximations is somehow cumbersome, however, we have identified those terms given the main contribution 
to the atom's quantum dynamics. In particular, for $|2m\omega/\omega_{0}| < 1$ the master equation is:
%%%%%%%%%%%%%%%%%%%%%%%%%%%%%%%%%%%%%%%%%%%%%%%%%%%%%%%%%%%%%%%%%%
\begin{eqnarray}
\frac{d}{dt}\bigl\langle Q(t)\bigr\rangle &-& \frac{i}{\hbar}\bigl\langle \bigl[\bar H_{0}, Q(t)\bigr]\bigr\rangle = 
-\gamma(t)\bigl\langle R^{+}\bigl[R^{-},Q(t)\bigr]\bigr\rangle \nonumber \\
&-& \gamma^{\ast}(t)\bigl\langle \bigl[Q(t),R^{+}\bigr]R^{-}\bigr\rangle. \label{MeqDD}
\end{eqnarray}
%%%%%%%%%%%%%%%%%%%%%%%%%%%%%%%%%%%%%%%%%%%%%%%%%%%%%%%%%%%%%%%%%%
Here,
%%%%%%%%%%%%%%%%%%%%%%%%%%%%%%%%%%%%%%%%%%%%%%%%%%%%%%%%%%%%%%%%%%
\begin{eqnarray*}
\gamma(t) &=& \frac{\gamma}{2}\sum^{\infty}_{n=-\infty}\sum^{\infty}_{m=-\infty}e^{2i(m-n)(\omega t + \phi)}\chi_{n}(x,\eta)\chi_{m}(x,\eta) 
\nonumber \\
&\times& \bigl(1+ x^{2}/4 + 2m\omega/\omega_{0}\bigr)^{3},
\end{eqnarray*}
%%%%%%%%%%%%%%%%%%%%%%%%%%%%%%%%%%%%%%%%%%%%%%%%%%%%%%%%%%%%%%%%%%
with $\gamma$ being the single-atom spontaneous decay rate at the bare transition frequency $\omega_{0}$, i.e. 
$\gamma = 4d^{2}\omega^{3}_{0}/(3\hbar c^{3})$, whereas 
%%%%%%%%%%%%%%%%%%%%%%%%%%%%%%%%%%%%%%%%%%%%%%%%%%%%%%%%%%%%%%%%%%
\begin{eqnarray*}
x=2\Omega/\omega_{0},
\end{eqnarray*}
and
\begin{eqnarray*}
\chi_{n}(x,\eta) &=& \bigl(1-x^{2}(1+n/\eta)/4\bigr)J_{n}(\eta).
\end{eqnarray*}
%%%%%%%%%%%%%%%%%%%%%%%%%%%%%%%%%%%%%%%%%%%%%%%%%%%%%%%%%%%%%%%%%%
Here we have used the relation: 
%%%%%%%%%%%%%%%%%%%%%%%%%%%%%%%%%%%%%%%%%%%%%%%%%%%%%%%%%%%%%%%%%%
\begin{eqnarray*}
J_{n-1}(\eta) + J_{n+1}(\eta) =2nJ_{n}(\eta)/\eta. 
\end{eqnarray*}
%%%%%%%%%%%%%%%%%%%%%%%%%%%%%%%%%%%%%%%%%%%%%%%%%%%%%%%%%%%%%%%%%%
Also, in the numerical simulations we shall truncate the summation range 
$(-\infty,\infty)$ to $(-n_{0},n_{0})$ such that for a selected value of $\eta$ one has $J_{n_{0}}(\eta) \to 0$ as well as $|2n_{0}\omega/\omega_{0}|<1$. 
Note that the spontaneous decay processes at the laser frequency $\omega$ are too small to influence the whole quantum dynamics and, therefore, 
are not taken into account. 
%%%%%%%%%%%%%%%%%%%%%%%%%%%%%%%%%%%%%%%%%%%%%%%%%%%%%%%%%%%%%%%%%%%
\begin{figure}[t]
\includegraphics[width=4.26cm]{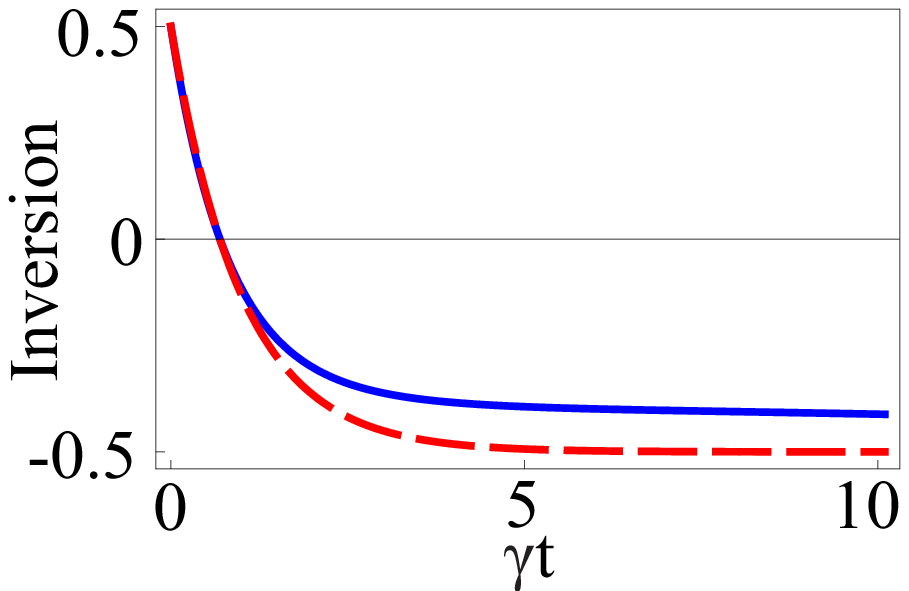}
\includegraphics[width=4.26cm]{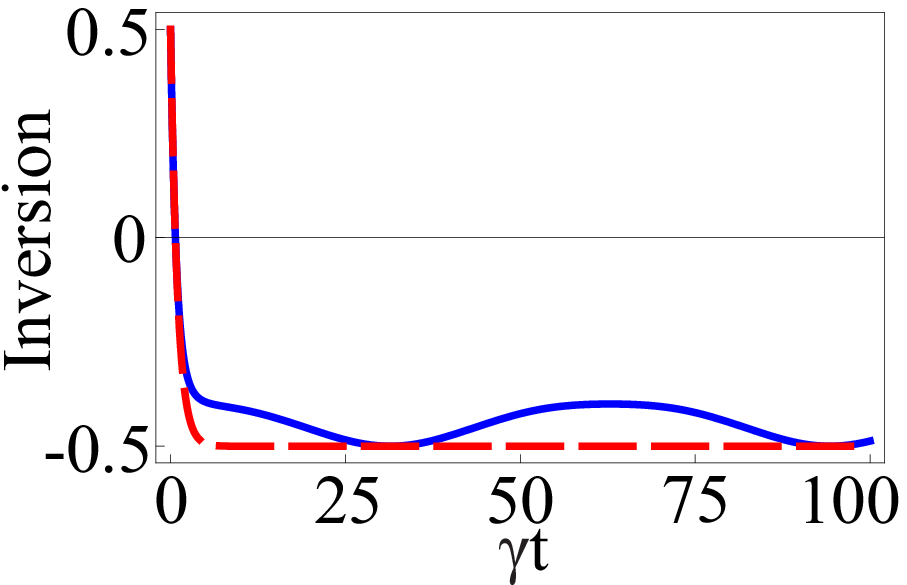} 
\begin{picture}(0,0)
\put(-20,75){(a)}
\put(100,75){(b)}
\end{picture}
\caption{\label{fig1} 
The spontaneous decay law, given by the mean-value of the inversion operator $\langle S_{z}(t)\rangle$, as a function of time in units of the inverse 
spontaneous decay rate at the bare transition frequency. Here $(2\Omega/\omega_{0})^{2}=0.64$, $\omega_{0}/\omega=2\cdot10^{4}$, 
$\omega/\gamma=0.05$ and $\phi=0$. The dashed line depicts the standard spontaneous decay dynamics of an excited two-level emitter in absence 
of any coherent driving.}
\end{figure}
%%%%%%%%%%%%%%%%%%%%%%%%%%%%%%%%%%%%%%%%%%%%%%%%%%%%%%%%%%%%%%%%%%%

%%%%%%%%%%%%%%%%%%%%%%%%%%%%%%%%%%%%%%%%%%%%%%%%%%%%%%%%%%%%%%%%%%
\section{Results and discussion}
%%%%%%%%%%%%%%%%%%%%%%%%%%%%%%%%%%%%%%%%%%%%%%%%%%%%%%%%%%%%%%%%%%
In the following, we shall describe the quantum dynamics of an excited two-level emitter interacting with a classical low-frequency 
and intense laser field based on transformation (\ref{rs}) and Eq.~(\ref{MeqDD}).
%%%%%%%%%%%%%%%%%%%%%%%%%%%%%%%%%%%%%%%%%%%%%%%%%%%%%%%%%%%%%%%%%%
\subsection{The case $\eta < 1$}
%%%%%%%%%%%%%%%%%%%%%%%%%%%%%%%%%%%%%%%%%%%%%%%%%%%%%%%%%%%%%%%%%%
Initially, we begin by investigating the spontaneous emission effect involving only few/several laser photon processes. This can be achieved when the 
parameter $\eta$ is smaller than unity. Let's consider, for instance, that $2\Omega/\omega_{0}=10^{-2}$ while $\omega_{0}/\omega=8\times 10^{3}$ 
then one has that $\eta=0.1$. Using the fact that $J_{n}(\eta) \approx \eta^{n}\bigl(1-\eta^{2}/[4(1+n)]\bigr)/(2^{n}n!)$ if $\eta \ll 1$, then for a 
$2n_{0}\omega$ process with $n_{0}=1$ one has that $\gamma(t) \approx \gamma\bigl(1 - (\eta^{2}/2)\cos{[4(\omega t+\phi)]}\bigr)/2$. Notice 
here that we have neglected the contributions smaller than $\gamma \eta^{2}$ in the total decay rate. Under this circumstance, the coherent evolution 
described by the Hamiltonian $\bar H_{0}$ plays no role and the spontaneous decay process of an excited two-level emitter in a low-frequency strong 
laser field is characterized by the usual exponential decay law, namely,
%%%%%%%%%%%%%%%%%%%%%%%%%%%%%%%%%%%%%%%%%%%%%%%%%%%%%%%%%%%%%%%%
\begin{eqnarray}
\langle S_{z}(t)\rangle &\approx& -1/2  + \exp\biggl[-2\int^{t}_{0}d\tau \gamma(\tau)\biggr] \nonumber \\
& \approx & -1/2  + \exp\bigl[-\gamma t\bigr].  \label {sz}
\end{eqnarray}
%%%%%%%%%%%%%%%%%%%%%%%%%%%%%%%%%%%%%%%%%%%%%%%%%%%%%%%%%%%%%%%%
The explanation for a $n_{0}=1$ spontaneous decay process is as follows: the decay channels at frequencies $\bar \omega_{0} \pm 2\omega$ and 
$\bar \omega_{0}$ lead to mutual cross-correlations such that the extra-induced decay channels cancel each other when $2\omega/\omega_{0} 
\ll \eta <1$. However, the cross-correlations among the channels $\bar \omega_{0} + 2\omega$ and $\bar \omega_{0} - 2\omega$ lead to a small 
oscillatory contribution, i.e. $\eta^{2}\cos{[4(\omega t + \phi)]}$, which does not affect the spontaneous decay. Generalizing in this way, even 
higher photon number processes, i.e. with $n_{0} >1$, do not modify the standard well-known exponential decay law as long as  
$2\Omega/\omega_{0} \ll \eta < 1$. 

%%%%%%%%%%%%%%%%%%%%%%%%%%%%%%%%%%%%%%%%%%%%%%%%%%%%%%%%%%%%%%%%
\subsection{The case $\eta \ge 1$ or $\eta \gg 1$}
%%%%%%%%%%%%%%%%%%%%%%%%%%%%%%%%%%%%%%%%%%%%%%%%%%%%%%%%%%%%%%%%
In this case, i.e. $\eta \ge 1$ or $\eta \gg 1$ with $2\Omega/\omega_{0} < 1$, the quantum dynamics of an excited two-level emitter interacting 
with a classical strong low-frequency laser field is determined by multi-photon processes. We have found that there is no deviation of the spontaneous 
decay from the standard one as long as $\eta \ge 1$. However, it is modified for $\eta \gg 1$ and larger values of $2\Omega/\omega_{0}$, with 
$2\Omega/\omega_{0}<1$. 

In Figure~(\ref{fig1}), we show the spontaneous decay law of an excited two-level emitter interacting with a low-frequency and strong classical 
coherent light source. The standard exponential quantum decay dynamics is clearly modified (compare the dashed and solid curves in Fig.~\ref{fig1}a). 
However, if one check longer time-durations then it can be seen that the decaying emitter starts following the applied field (see Fig.~\ref{fig1}b).
It looks like we have an interplay among the exponential spontaneous decay and incomplete Rabi oscillations due to the low-frequency 
coherent driving field. Nevertheless, one can still have a modification of the exponential spontaneous decay because of the quantum interference 
processes among the induced decay channels. We will return to this issue later. Also, importantly, for $(2\Omega/\omega_{0})^{2}=0.64$ as it is 
the case in Figure~(\ref{fig1}), we have considered expansions terms up to $(2\Omega/\omega_{0})^{8}$ in expression (\ref{omb}). In this case, 
the time-dependent spontaneous decay rate in Eq.~(\ref{MeqDD}) is given by the following expression
%%%%%%%%%%%%%%%%%%%%%%%%%%%%%%%%%%%%%%%%%%%%%%%%%%%%%%%%%%%%%%%%%%
\begin{widetext}
%%%%%%%%%%%%%%%%%%%%%%%%%%%%%%%%%%%%%%%%%%%%%%%%%%%%%%%%%%%%%%%%%%
\begin{eqnarray}
\gamma(t) &=& \frac{\gamma}{2}\sum_{n,n'}\sum_{m,m'}\sum_{s,s'}\sum_{r,r'}e^{2i(n-n')\phi(t)}e^{-4i(m-m')\phi(t)}e^{6i(s-s')\phi(t)}e^{-8i(r-r')\phi(t)}
\biggl(1 +  x^{2}/4  - 3x^{4}/64 + 5x^{6}/256 \nonumber \\
&-& 175x^{8}/16384 + 2\bigl(n-2m+3s-4r\bigr)\omega/\omega_{0}\biggr)^{3}\chi_{nmsr}\bigl(x,\bar \eta,\bar \xi,\bar \beta,\rho\bigr)
\chi_{n'm's'r'}\bigl(x,\bar \eta,\bar \xi,\bar \beta,\rho\bigr),
\label{gmn}
\end{eqnarray}
%%%%%%%%%%%%%%%%%%%%%%%%%%%%%%%%%%%%%%%%%%%%%%%%%%%%%%%%%%%%%%%%%%
%%%%%%%%%%%%%%%%%%%%%%%%%%%%%%%%%%%%%%%%%%%%%%%%%%%%%%%%%%%%%%%%%%
\end{widetext}
%%%%%%%%%%%%%%%%%%%%%%%%%%%%%%%%%%%%%%%%%%%%%%%%%%%%%%%%%%%%%%%%%%
where we have assumed that $|2(n-2m+3s-4r)\omega/\omega_{0}| < 1$, whereas
%%%%%%%%%%%%%%%%%%%%%%%%%%%%%%%%%%%%%%%%%%%%%%%%%%%%%%%%%%%%%%%%%%
\begin{eqnarray*}
\bar H_{0} &=& i\hbar\bar \alpha(t)\sum_{n,m,s,r}J_{n}(\bar \eta)J_{m}(\bar \xi)J_{s}(\bar \beta)J_{r}(\rho)e^{-2i(n-2m)\phi(t)} \nonumber \\
&\times&e^{-i(\tilde \omega_{0}t - \bar \eta\sin{2\phi} + \bar \xi\sin{4\phi}-\bar \beta\sin{6\phi}+\rho\sin{8\phi})} \nonumber \\
&\times& e^{-2i(3s-4r)\phi(t)}R^{-} + H.c..
\end{eqnarray*}
%%%%%%%%%%%%%%%%%%%%%%%%%%%%%%%%%%%%%%%%%%%%%%%%%%%%%%%%%%%%%%%%%%
Here $\chi_{nmsr}(x,\bar \eta,\bar \xi,\bar \beta,\rho) =J_{n}(\bar \eta)J_{m}(\bar \xi)J_{s}(\bar \beta)J_{r}(\rho)
\bigl\{1-x^{2}/4+9x^{4}/64 - 25x^{6}/256+35^{2}x^{8}/128^{2}-nx^{2}(1-3x^{2}/4+75x^{4}/128-245x^{6}/512)/(4\bar \eta) + 
3mx^{4}(1-5x^{2}/4+245x^{4}/192)/(64\bar \xi) - 5sx^{6}(1-7x^{2}/4)/(512\bar \beta)+35rx^{8}/(128^{2}\rho)\bigr\}$, with 
%%%%%%%%%%%%%%%%%%%%%%%%%%%%%%%%%%%%%%%%%%%%%%%%%%%%%%%%%%%%%%%%%%
\begin{eqnarray*}
\bar \eta &=& \eta\bigl(1-x^{2}/4+15x^{4}/128-35x^{6}/512\bigr),  \\
\bar \xi &=& \xi\bigl(1-3x^{2}/4+35x^{4}/64\bigr),  \\
\bar \beta &=& \beta\bigl(1-5x^{2}/4\bigr),
\end{eqnarray*}
%%%%%%%%%%%%%%%%%%%%%%%%%%%%%%%%%%%%%%%%%%%%%%%%%%%%%%%%%%%%%%%%%%
whereas $\xi=(\omega_{0}/\omega)(x/4)^{4}$, $\beta=(\omega_{0}/\omega)x^{6}/3072$, and $\rho=10(\omega_{0}/\omega)x^{8}/8^{6}$. 
Further, $\bar \alpha(t) \approx x\omega\sin{\phi(t)}\bigl(1 - x^{2}\cos^{2}{\phi(t)}+x^{4}\cos^{4}{\phi(t)}-x^{6}\cos^{6}\phi(t)\bigr)/2$,  
with $\phi(t)=\omega t + \phi$, while
%%%%%%%%%%%%%%%%%%%%%%%%%%%%%%%%%%%%%%%%%%%%%%%%%%%%%%%%%%%%%%%%%%
\begin{eqnarray}
\tilde \omega_{0}=\omega_{0}\biggl(1 + \frac{x^{2}}{4} - \frac{3x^{4}}{64} + \frac{5x^{6}}{256} - \frac{175x^{8}}{16384}\biggr). \label{omm}
\end{eqnarray}
%%%%%%%%%%%%%%%%%%%%%%%%%%%%%%%%%%%%%%%%%%%%%%%%%%%%%%%%%%%%%%%%%%
Note that while we restricted the expansion of Exp.~(\ref{omb}) to a certain order in $x$, in the subsequent calculations we did not. Respectively, one 
can obtain the time-dependent decay rates for additional expansion terms in Exp.~(\ref{omb}). Generally, these decay rates will be proportional with a 
product of Bessel functions. We have observed that when the argument of one of the Bessel function is much smaller than unity then the spontaneous 
quantum dynamics does not change if one add further expansion terms in (\ref{omb}). Moreover, the modification of the spontaneous decay law is more 
pronounced for larger values of the ratio $2\Omega/\omega_{0} < 1$.  As a real system, where this prediction can be checked, may be considered 
certain solid state media \cite{kal}. Higher decay rates, $\gamma \sim 10^{12}{\rm Hz}$, at transition frequencies $\omega_{0} \sim 10^{15}{\rm Hz}$ 
are proper to such systems. Therefore, for $\omega_{0}/\omega \sim 2\cdot 10^{4}$ one has $\omega/\gamma \sim 0.05$. In Figure~(\ref{fig1}) the 
Rabi frequency's value corresponds to $\Omega \sim 4 \times 10^{14}{\rm Hz}$. In this case, a transition dipole moment 
$d \sim 2\times 10^{-29}{\rm C\cdot m}$ would lead to an electric field amplitude of the order of $E_{L} \sim 10^{9}{\rm V/m}$. The ionization processes 
can be avoided if the ionization time, $t_{i}$, is larger than $t_{i} > 10^{-11}{\rm s}$.
%%%%%%%%%%%%%%%%%%%%%%%%%%%%%%%%%%%%%%%%%%%%%%%%%%%%%%%%%%%%%%%%%%%
\begin{figure}[b]
\includegraphics[width=7.5cm]{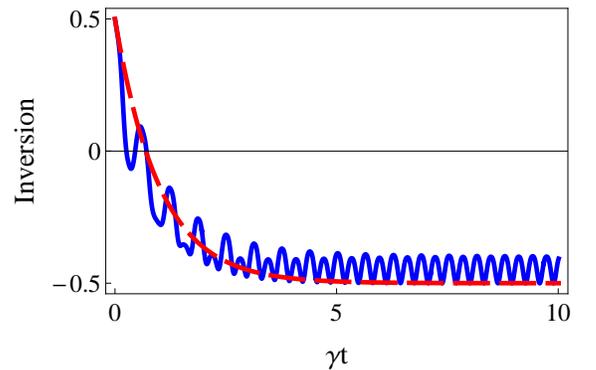}
\caption{\label{fig2} The same as in Figure~{\ref{fig1}}(a) but for $\omega/\gamma=10$.}
\end{figure}
%%%%%%%%%%%%%%%%%%%%%%%%%%%%%%%%%%%%%%%%%%%%%%%%%%%%%%%%%%%%%%%%%%%

We turn further to Figure  (\ref{fig2}) where we show the spontaneous quantum dynamics when the laser frequency is larger than the spontaneous 
decay rate. At the beginning of the evolution there is a fast population decay which is identified with the strong low-frequency driving rather than to 
quantum interference effects. Consequently, once the emitter decays to the ground state it will oscillate, in the ground state, due to strong  continuous 
coherent wave driving. 
%%%%%%%%%%%%%%%%%%%%%%%%%%%%%%%%%%%%%%%%%%%%%%%%%%%%%%%%%%%%%%%%%%%
\begin{figure}[t]
\includegraphics[width=4.26cm]{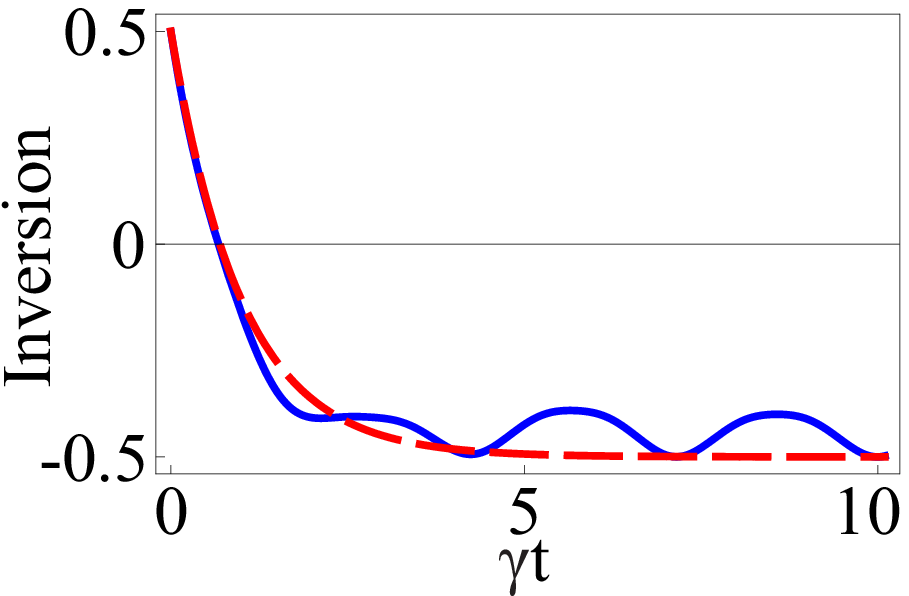}
\includegraphics[width=4.26cm]{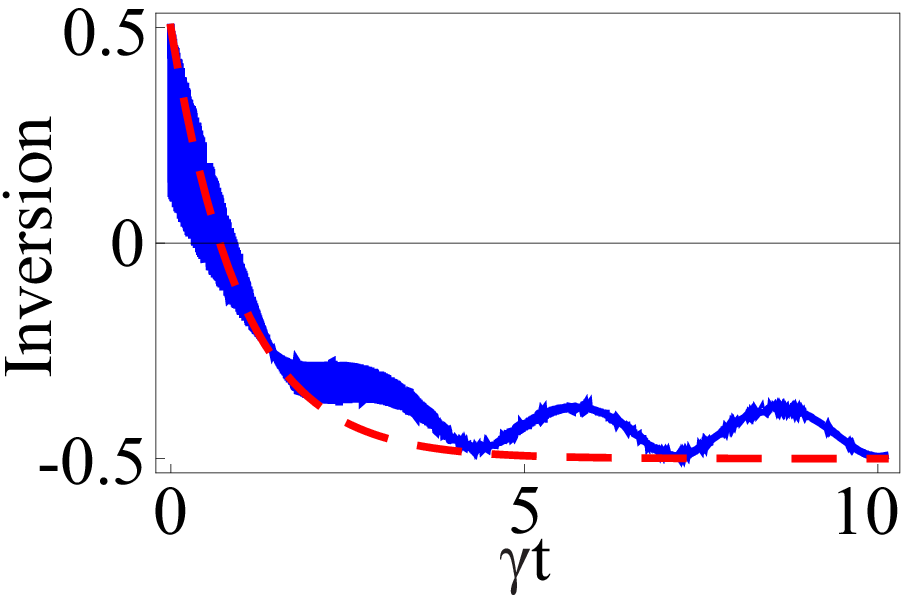}
\begin{picture}(0,0)
\put(-20,75){(a)}
\put(105,75){(b)}
\end{picture}
\caption{\label{fig3} 
(a) The population inversion $\langle S_{z}(t)\rangle$ as a function of $\gamma t$ obtained with the analytical approach developed here, while 
$\omega/\gamma=1.1$. (b) The same obtained from the master equation (\ref{sme}). Other parameters are as in Figure~(\ref{fig1}).}
\end{figure}
%%%%%%%%%%%%%%%%%%%%%%%%%%%%%%%%%%%%%%%%%%%%%%%%%%%%%%%%%%%%%%%%%%%
%%%%%%%%%%%%%%%%%%%%%%%%%%%%%%%%%%%%%%%%%%%%%%%%%%%%%%%%%%%%%%%%%%%
\begin{figure}[b]
\includegraphics[width=7cm]{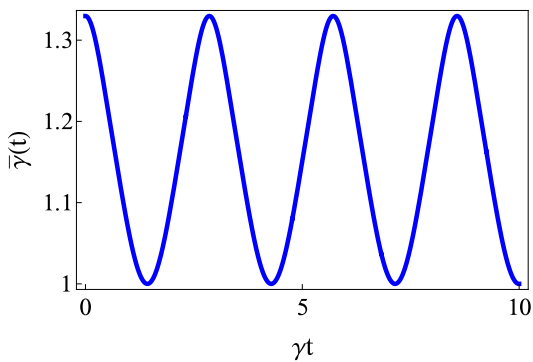}
\caption{\label{fig4} The time-dependent decay rate, i.e. $\bar \gamma(t)=\bigl(\gamma(t)+\gamma(t)^{\ast}\bigr)$ [in units of $\gamma$] 
evaluated with the help of the expression (\ref{gmn}) as a function of $\gamma t$, for $\omega/\gamma=1.1$. Other parameters are as in 
Figure~\ref{fig3}(a).}
\end{figure}
%%%%%%%%%%%%%%%%%%%%%%%%%%%%%%%%%%%%%%%%%%%%%%%%%%%%%%%%%%%%%%%%%%%

To additionally prove our conclusion, in what follows, we compare our results with those obtained with a standard master equation where the spontaneous 
emission is introduced in the usual way \cite{ag,al,sczb,gxl,fc,rew}, namely,
%%%%%%%%%%%%%%%%%%%%%%%%%%%%%%%%%%%%%%%%%%%%%%%%%%%%%%%%%%%%%%%%%%%
\begin{eqnarray}
\frac{d}{dt}\bigl\langle Q(t)\bigr\rangle &=&i\bigl\langle \bigl[\omega_{0}S_{z} - \Omega\cos{\bigl(\omega t + \phi\bigr)}\bigl(S^{+}+S^{-}\bigr),Q\bigr]
\bigr\rangle \nonumber \\
&-&\frac{\gamma}{2}\bigl(\bigl\langle S^{+}\bigl[S^{-},Q\bigr]\bigr\rangle + \bigl\langle\bigl[Q, S^{+}\bigr]S^{-}\bigr\rangle \bigr). \label{sme}
\end{eqnarray}
%%%%%%%%%%%%%%%%%%%%%%%%%%%%%%%%%%%%%%%%%%%%%%%%%%%%%%%%%%%%%%%%%%%
We have found that as long as $\omega/\gamma \ll 1$ the results obtained with the analytical formalism described here and the master equation 
(\ref{sme}) looks somehow similar. This fact does not infirm the existence of quantum interference effects. The reason is that in our approach, 
due to strong laser-pumping, the transition frequency is increased by $10$ percent when $x=0.8$, see expression~(\ref{omm}), meaning that 
the spontaneous decay should be faster. However, we obtain almost the same results as those obtained with the master equation (\ref{sme}). 
This means that the spontaneous decay was slowed down and this is the reason of the correspondence with the master equation (\ref{sme}) 
which does not contain the modification of the transition frequency due to strong pumping nor various induced decay channels. When the frequency 
of the applied field is of the order of the bare spontaneous decay rate we observe slightly different behaviors, see Figure~(\ref{fig3}). The initial 
time evolution is faster than the standard exponential spontaneous decay law, when it is described by our formalism and, thus, quantum interference 
is responsible for the rapid decay evolution. In this context, Figure~(\ref{fig4}) depicts the time dependence of the scaled decay rate 
$\bar \gamma(t)/\gamma \equiv \bigl(\gamma(t) + \gamma(t)^{\ast}\bigr)/\gamma$ given by the Exp.~(\ref{gmn}). A time dependent decay rate 
presented here may help to understand the spontaneous emission dynamics of the excited emitter (although it will enter in that dynamics integrated, 
see for instance, the first line of Eq.~\ref{sz}). The fact that the magnitude of the decay rate is larger than the single-qubit bare decay rate is due 
to the frequency shift, see expression~(\ref{omm}), arising from the strongly applied low-frequency coherent driving, i.e., the external field do modify 
it. Also, when $\bar \gamma(t)/\gamma \approx 1$ the spontaneous decay is faster than the usual single-qubit spontaneous decay law obtained in the 
absence of any coherent pumping, compare Figures (\ref{fig3}a) and (4), respectively. Notice here that the reference time, i.e. $t=0$, is taken at 
$t \sim \Omega^{-1}$, i.e. we have performed the secular approximation. Generalizing in this way, the spontaneous emission is modified because of 
an interplay among slow classical and strong coherent pumping wave and additionally induced spontaneous interfering decay channels. 

Finally, we note that there is a substantial progress towards control of the spontaneous emission processes. Most of the studies use either near resonant driving or 
strong low-frequency quantized or classical applied fields \cite{fc,rew}. In the latter case, the spontaneous emission inhibition occurs via additional energy 
levels or/and modification of the environmental vacuum reservoir, and based on markovian or non-markovian processes \cite{ek1,ek2,shap,kur}. In the 
present study, however, we focused on an isolated two-level qubit pumped by a strong and low-frequency coherent field, without auxiliary off-resonant atomic states, and coupled to the regular electromagnetic vacuum modes. We find that the spontaneous emission modification is not too drastic, which in part is due to the fact that only the driving field properties remain as control parmeters in our scheme. But comparing this result to those of the model in~\cite{ek1,ek2}, in which  the low-frequency field can induce interfering multiphoton decay pathways via additional off-resonant auxiliary energy levels,  we may further conclude that these additional multiphoton decay pathways are crucial for the strong spontaneous emission modification found there.

%%%%%%%%%%%%%%%%%%%%%%%%%%%%%%%%%%%%%%%%%%%%%%%%%%%%%%%%%%%%%%%%%%

%%%%%%%%%%%%%%%%%%%%%%%%%%%%%%%%%%%%%%%%%%%%%%%%%%%%%%%%%%%%%%%%%%
\section{Summary}
%%%%%%%%%%%%%%%%%%%%%%%%%%%%%%%%%%%%%%%%%%%%%%%%%%%%%%%%%%%%%%%%%%
We have investigated the interaction of an excited two-level emitter with a coherent and strong low-frequency classical electromagnetic field. More precisely, we were interested in the quantum dynamics of the spontaneous emission processes. We have found that the spontaneous emission decay of an initially excited atom is slowed down or accelerated via the action of a strong and coherent classical low-frequency electromagnetic wave. The reasons are the presence of external low-frequency pumping followed by additionally induced decay channels that lead to destructive or constructive quantum interference phenomena and, consequently, to modification of the spontaneous emission. Furthermore, the induced spontaneous decay processes involve an even laser-photon number. Also, the modification of the bare transition frequency due to the strong low-frequency applied field is shown as well. An interesting perspective is to extend the present or related analysis on the effect of intense low-frequency fields beyond atoms, e.g., involving molecules
driven by resonant low-frequency laser radiation~\cite{jetp3}, or multi-photon processes in artificial quantum systems like superconducting quantum circuits \cite{art,art1,art2}, quantum dot \cite{art3,art4}  or off-resonantly driven solid-state spin systems \cite{art7}. This way, more versatile parameter ranges may become possible.

%%%%%%%%%%%%%%%%%%%%%%%%%%%%%%%%%%%%%%%%%%%%%%%%%%%%%%%%%

%%%%%%%%%%%%%%%%%%%%%%%%%%%%%%%%%%%%%%%%%%%%%%%%%%%%%%%%%%%%%%%%%%%
\acknowledgments
%%%%%%%%%%%%%%%%%%%%%%%%%%%%%%%%%%%%%%%%%%%%%%%%%%%%%%%%%%%%%%%%%%%
M.M. is grateful for the nice hospitality of the Theory Division of the Max Planck Institute for Nuclear Physics from Heidelberg, Germany. 
Furthermore, M.M. and C.H.K. acknowledge the financial support by the German Federal Ministry of Education and Research, grant 
No. 01DK13015, and the Moldavian National Agency for Research and Development, grants No. 13.820.05.07/GF and 20.80009.5007.07, 
respectively. 
%%%%%%%%%%%%%%%%%%%%%%%%%%%%%%%%%%%%%%%%%%%%%%%%%%%%%%%%%

%%%%%%%%%%%%%%%%%%%%%%%%%%%%%%%%%%%%%%%%%%%%%%%%%%%%%%%%%%%%%%%%%%

%%%%%%%%%%%%%%%%%%%%%%%%%%%%%%%%%%%%%%%%%%%%%%%%%%%%%%%%%%%%%%%%
%%%%%%%%%%%%%%%%%%%%%%%%%%%%%%%%%%%%%%%%%%%%%%%%%%%%%%%%%%%%%%%%
\end{document}
%%%%%%%%%%%%%%%%%%%%%%%%%%%%%%%%%%%%%%%%%%%%%%%%%%%%%%%%%%%%%%%%